\documentclass[aps,prl,twocolumn,superscriptaddress]{revtex4}
\usepackage{amsmath}
\usepackage{amssymb}
\usepackage{graphicx}
\usepackage{dcolumn}
\usepackage{bm}
\makeatletter
\def\captionof#1#2{{\def\@captype{#1}#2}}
\makeatother

\begin{document}

\title{Quantum Non-Equilibrium Steady States Induced by Repeated Interactions}
\author{Dragi Karevski}
\email{karevski@lpm.u-nancy.fr}
\affiliation{Institut Jean Lamour, dpt. P2M, Groupe de Physique Statistique, Nancy-Universit\'e CNRS,
B.P. 70239, F-54506 Vandoeuvre les Nancy Cedex, France}


\author{Thierry Platini}
\affiliation{Department of Physics, Virginia Tech, Blacksburg, VA 24061, USA}

\begin{abstract}
We study the steady state of a finite XX chain coupled at its boundaries to 
quantum reservoirs made of free spins that interact one after the other with the chain. The two-point correlations are calculated exactly and it is shown that the steady state is completely characterized by the magnetization profile and the associated current. Except at the boundary sites, the magnetization is given by the average of the reservoirs' magnetizations. The steady state current, proportional to the difference in the reservoirs' magnetizations, shows a non-monotonous behavior with respect to the system-reservoir coupling strength, with an optimal current state for a finite value of the coupling. Moreover, we show that the steady state can be described by a generalized Gibbs state. 
\end{abstract}

\pacs{Valid PACS appear here}
\maketitle

Understanding non-equilibrium behavior of quantum systems on the basis of general principles is one of the more challenging prospects of statistical physics. In particular, the complete characterization of the so-called Quantum Non-Equilibrium Steady-States (QNESS), {\it i. e.}  stationary current full quantum states, is of primer and central focus since they are possible candidates for playing a role similar to Gibbs states in constructing a non-equilibrium statistical mechanics \cite{landauer,buttiker,kubo,zubarev,maclennan,hershfield,gallavotti,komatsu}. 

To elucidate the general guiding principles for a non-equilibrium statistical mechanics, exactly solvable models play a central role. Among many models, the XX quantum chain is one of the simplest non-trivial many-body system. 
Its $N$-sites Hamiltonian is given by
\begin{equation}
H_S=-\frac{J}{2}\sum_{l=1}^{N-1} [\sigma_l^x\sigma_{l+1}^x+\sigma_l^y\sigma_{l+1}^y]+\frac{h}{2}\sum_{l=1}^N \sigma^z_l
\label{XX1}
\end{equation}
where the $\sigma$s are the usual Pauli matrices, $J$ is the exchange coupling and $h$ a transverse (possibly external) magnetic field.
Due to the fact that its dynamics can be described in an explicit way, the one-dimensional XX model has been extensively studied in various non-equilibrium contexts \cite{karevski1}.
On the experimental side, the most promising perspectives for these studies come from the ultra-cold atoms community since the XX model (\ref{XX1}) can be mapped on a one-dimensional Hard-Core boson (Tonks-Girardeau \cite{girardeau}) model
through the transformation $b_l^{+}=(\sigma_l^x+ i \sigma_l^y)/2$ and $b_l=(\sigma_l^x- i \sigma_l^y)/2$.
Experiments on such 1D hard-core bosons have been performed with Rubidium atoms in both continuum~\cite{kinoshita1} and lattice \cite{paredes} versions.

Antal {\it et al.} \cite{antal} studied the ground state of the Ising and XX chain Hamiltonian with the addition of a magnetization or energy current $\cal J$ via a Lagrange multiplier. The ground state of the effective Hamiltonian $H_S-\lambda {\cal J}$  was interpreted as a non-equilibrium stationary current full state. Such an effective Hamiltonian was supposed to capture locally the essential features of a finite chain coupled at its boundary sites to quantum reservoirs.
Soon after, Ogata \cite{ogata}, Aschbacher and Pillet \cite{asch} considered the anisotropic XY steady state induced by the unitary dynamics, $U(t)=e^{-itH_S}$, starting with an initial state in which the left and right halves are set at inverse temperature $\beta_L$ and $\beta_R$. They showed that the QNESS can be effectively described by a generalized Gibbs state $\sim e^{-\bar{\beta}H_S+\delta Y}$ where  $\bar{\beta}=(\beta_L+\beta_R)/2$ is the average inverse temperature, $\delta=(\beta_L-\beta_R)/2$ is a driving force coupled to a long-range operator $Y$ commuting with $H_S$. The operator $Y$ is given by $Y=\sum_{l=1}^{\infty} \mu_l Y_l$ where $Y_l$ are currents operators associated to $l$th sites conserved quantities and where the coefficients $\mu_l$ show a power law decay ( $\sim 1/l$). 
The long range of $Y$ is a signature of a strong nonlocal properties of the QNESS which is believed to be a generic feature of NESS \cite{ogata}. One may notice that the Antal {\it et al.} steady state relates to the same effective Hamiltonian  truncating the current series $Y=\sum_l \mu_l Y_l$ to the first few terms.


In this work we study the $N$-site isotropic XX-chain (\ref{XX1}) coupled at its boundary sites to quantum reservoirs at different temperatures.  In the anisotropic case a recent study has been reported in \cite{prosen} with Markovian baths. Here,
the left and right reservoirs are made of an infinite set of non-interacting spins 1/2  in the same transverse field $h$ with Hamiltonian 
$
H_E^{L(R)}=\sum_{n=1}^{\infty}H_n^{L(R)}=h\sum_{n=1}^{\infty}{b^{L(R)}_n}^+b^{L(R)}_n
$.
The system-reservoir couplings are implemented via a repeated interaction scheme, meaning that each subsystem (particle) composing the reservoirs are interacting with the system one after the other \cite{AttalPautrat}. 
To have a physical picture of such an interaction scheme, one may think of a laser (or particle) beam falling on the system.  The system-reservoir interaction is given by the time-dependent Hamiltonian $V(t)=V_L(t)+V_R(t)$ where $V_{L(R)}(t)=V_{L(R)}^{n}$ $\forall t\in ](n-1)\tau,n\tau]$ with
\begin{equation}
V_{L(R)}^{n}=-J_{E}[{b_n^{L(R)}}^+b_{1(N)}+b^+_{1(N)}{b_n^{L(R)}}]\; ,
\end{equation}
selecting the $n$th left and right reservoir spins
in the time-interval $t\in ](n-1)\tau,n\tau]$. 

We start at $t=0$ with a system-environment decoupled initial state $\rho(0)=\rho_S(0)\otimes\rho_E(0)$, where $\rho_S$ is an equilibrium state of the system and where the environment density matrix is given by $\rho_E=\otimes_{\mathbb{N}^*}\rho_n=\otimes_{\mathbb{N}^*}(\rho_n^L\otimes\rho_n^R)$ with  one-particle thermal density matrices 
$
\rho^{L(R)}_n=\frac{1}{Z_{L(R)}}e^{-\beta_{L(R)}h{b^{L(R)}_n}^+b^{L(R)}_n}=
\frac{1+m_{L(R)}}{2}|+\rangle\langle +|+
\frac{1-m_{L(R)}}{2}|-\rangle\langle -|
$ with $\sigma^z|\pm\rangle = \pm |\pm\rangle$ 
\cite{remark0}.
At $t=0^+$, the first left and right reservoir spins start to interact with the left and right system boundary spins for a time $\tau$ through the hopping term $V_{L(R)}^1$. 
At $t=\tau^+$, the first reservoir spins are replaced by the second ones interacting with the system through
$V_{L(R)}^{2}$ for a time $\tau$. The process is then repeated again and again \cite{remark1}. 
\begin{figure}
\centerline{
\includegraphics[width=7.5cm,angle=0]{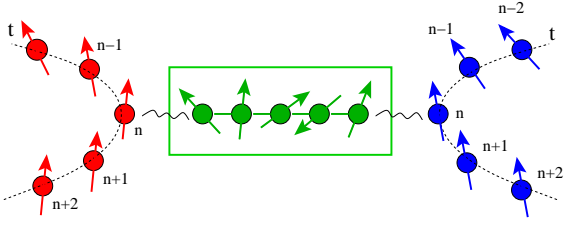}
}
\caption{\label{fig1} Sketch of a system in contact at both ends with  repeated interaction reservoirs.
}
\end{figure}
Iterating the process $n$-times, the reduced density matrix associated to the system part  is given at time $t=n\tau$ by
\begin{eqnarray}
\rho_S(n\tau)&=&Tr_E\left\{e^{-i\tau H_{Tot}^{\{n\}}}\rho((n-1)\tau)e^{i\tau H_{Tot}^{\{n\}}}
 \right\}\nonumber \\
 &=&Tr_n\left\{U_{I}^{\{n\}} \rho_S((n-1)\tau)\otimes \rho_n {U_{I}^{\{n\}}}^\dagger\right\}
\label{rho1}
\end{eqnarray}
where $H_{Tot}^{\{n\}}=H_S+H_E+V_I^{\{n\}}$ is the total Hamiltonian in the time-interval $[(n-1)\tau,n\tau]$ with the interaction part $V_I^{(n)}=V_L^{(n)}+V_R^{(n)}$, and where $U_{I}^{\{n\}}=e^{-i\tau (H_S+V_I^{\{n\}}+H_n)}$ with $H_n=H_n^L+H_n^R$.

To solve the recursive dynamical equation (\ref{rho1}), we introduce the fermionic representation of the coupled part of the total Hamiltonian, $H^{\{Sn\}}=H_S+V_I^{\{n\}}+H_n$,  which is of the form (\ref{XX1}) with $N+2$ sites, via the usual Jordan-Wigner transformation \cite{lieb}:
\begin{equation}
\begin{array}{rl}
{\Gamma}_k^1&=e^{i\pi\sum_{j=0}^{k-1}b^+_jb_j}(b_k+b_k^+)\\
{\Gamma}_k^2&=-ie^{i\pi\sum_{j=0}^{k-1}b^+_jb_j}(b_k-b_k^+)
\end{array}
\label{JW}
\end{equation}
where the $\Gamma$s are Majorana real (Clifford) operators satisfying ${\Gamma}^\dagger=\Gamma$ and $\{{\Gamma}_i^\alpha,{\Gamma}_j^\beta\}=2\delta_{ij}\delta_{\alpha\beta}$.
Notice here that the zeroth label is associated to the left-reservoir spin and the $N+1$th to the right one, keeping the $1,..,N$ labels for the system.
The interacting part of the total Hamiltonian takes the form 
$
H^{\{Sn\}}=1/4 {{\bf \Gamma}}^\dagger T_{Sn} {\bf \Gamma}
$
where ${{\bf \Gamma}}^\dagger =({\Gamma}_0^1,{\Gamma}_1^1,...,{\Gamma}_{N+1}^1,{\Gamma}_0^2,...,{\Gamma}_{N+1}^2)$
is a $2(N+2)$-component operator and where the $2(N+2)\times 2(N+2)$ matrix $T$ is given by
$
T_{Sn}=\left(\begin{array}{cc}
0&C_{Sn}\\
C_{Sn}^\dagger&0
\end{array}
\right)
$
with the tridiagonal matrix $\left(C_{Sn}\right)_{lm}=-i(h\delta_{lm}+J_l\delta_{lm-1}+J_m\delta_{lm+1})$ containing system-environment couplings $J_0=J_N=J_E$ and $J_l=J$ $\forall l=1,...,N-1$ for the system part.
The interesting point in using the $\Gamma$s is that 
their time evolution, generated by $H^{\{Sn\}}$, is simply given by a rotation $R(t)$: ${\bf \Gamma}(t)= { R}(t){\bf \Gamma}(0)=e^{-itT_{Sn}}{\bf \Gamma}(0)$ \cite{karevski1}. 

Since the total initial state is Gaussian in terms of fermions, the reduced system density matrix remains Gaussian during the repeated-interaction process \cite{peschel,cheong}. No many-body interactions are generated during the time evolution. 
As a consequence, thanks to Wick's theorem, one may characterize completely the state of the system at any time by its two-point correlation matrix $G_S(t)$ defined by 
$
\left(G_S(t)\right)_{jk}=\frac{i}{2}Tr_S\left\{ \big[({\bf \Gamma}_S)_k,({\bf \Gamma}_S)_j\big]\rho_S(t)\right\}
$. 
Along the same lines, the  two-point
correlation matrix $G^{\{Sn\}}(n\tau)$  characterizes completely at time $n\tau$ the total state of the $n$th environment copy + system. For these reasons, instead of computing directly the system-density matrix,
thanks to (\ref{rho1}), we will study the correlation matrix and reconstruct afterwards from it the density matrix.

Ordering the ${{\bf \Gamma}}^\dagger=({\bf \Gamma}_E^\dagger,{\bf \Gamma}_S^\dagger)$ such that the first part, ${\bf  \Gamma}_E^\dagger=({\bf  \Gamma}_L^\dagger,{\bf  \Gamma}_R^\dagger)$, is associated to the $n$th copy of the environment and that ${\bf \Gamma}_S^\dagger$ contains the components of the system, we write the rotation matrix $R(\tau)$  as
\begin{equation}
R(\tau)=e^{-i \tau T_{Sn}}
=\left(\begin{array}{cc}
R_E&R_{ES}\\
R_{SE}&R_S
\end{array}\right)\;.
\end{equation}
Using this decomposition into (\ref{rho1}), 
one arrives at the fundamental dynamical equation governing the system: 
\begin{equation}
G_S(n\tau)={ R}_S G_S((n-1)\tau){{ R}_S}^\dagger+{ R}_{SE} G_E{{ R}_{SE}}^\dagger\; .
\label{eqfond}
\end{equation}
The $4\times 4$ environment correlation matrix $G_E$ describes the initial environment state. It is evaluated with respect to the environment two-spin initial state $\rho_n^L\otimes\rho_n^R$. If the system is decoupled from the reservoirs ($J_E=0$)
the rotation matrix splits into a block diagonal form, with $R_{ES}=0$, $R_{SE}=0$,
reflecting the separate unitary evolution of the system and the environment through $R_S$ and  $R_E$ respectively.
Introducing the infinitesimal generator $\cal L$ through the dynamical map
$
\alpha_S^\tau(X)\equiv e^{-\tau {\cal L}}(X)\equiv R_S X {R_S}^{\dagger}
$
one may iterate equation (\ref{eqfond}) which becomes in the continuum limit
$
G_S(t)=e^{-t{\cal L}}\left(G_S(0)\right)+\int_0^t \!ds\; e^{-s{\cal L}}\left( \tilde{G}_E\right)
\label{eqcont}
$
with 
$
\tilde{G}_E\equiv R_{SE} G_E R_{SE}^{\dagger}
$.
The explicit form of the generator $\cal L$ depends on the proper way one rescales the interaction couplings in the continuum limit $\tau \rightarrow 0$. 
To take into account the non-trivial effect of the interaction between the system and the environment, one has to rescale the interaction couplings $J_E\rightarrow J_E/\sqrt{\tau}$. Other rescalings give either trivial limits or no limit at all \cite{AttalPautrat,remark2}.  The total $T$-matrix takes then the form
$
T_{Sn}=\left(\begin{array}{cc}
T_E&\Theta/\sqrt{\tau}\\
\Theta^\dagger/\sqrt{\tau}&T_S
\end{array}
\right)\; ,
$
where $T_E$ and $T_S$ are $T$-matrices of the environment and system parts respectively while $\Theta$ is a $4\times 2N$ matrix describing the system-environment interaction with components $(\Theta)_{kl}=-iJ_E (\delta_{k1}\delta_{lN+1}-\delta_{k2}\delta_{l1}+\delta_{k3}\delta_{l2N}-\delta_{k4}\delta_{lN})$.
Developing the exponential $R(\tau)=e^{-i\tau T_{Sn}}$ to the lowest order in $\tau$, and projecting to the system part leads to the Linblad-like differential equation 
$
\partial_t G_S(t)=-{\cal L}(G_S(t))+\Theta^\dagger G_E\Theta
\label{eqfondcont}
$
with the generator ${\cal L}(.)=i[T_S,.]+\frac{1}{2}\{\Theta^\dagger\Theta,.\}$. 

Using the expressions of the interaction matrix $T_{Sn}$ and the antisymmetry of the correlation matrix, one arrives finally at two coupled $N\times N$ matrix equations:
\begin{equation}\left\{\begin{array}{rl}
\partial_t G_{d}&=-i[G_{o},C_S]-\frac{J_E^2}{2}\Lambda_{d}\\
\partial_t G_{o}&=i[G_{d},C_S]-\frac{J_E^2}{2}\Lambda_{o}-J_E^2 M_E
\end{array}\right.
\label{eqfond3}
\end{equation}
where the $N\times N$ matrices  $G_{d}$ and $G_{o}$ are defined through 
$
G_S=\left(\begin{array}{cc}
G_{d}&G_{o}\\
-G_{o}^{T}&G_{d}
\end{array}
\right)
$ and where $C_S$ is the restriction of $C_{Sn}$ to the system part. 
The matrix $M_E$ with elements
$(M_E)_{kl}=m_L\delta_{k1}\delta_{l1}+m_R\delta_{kN}\delta_{lN}$ contains the left and right environment magnetizations. The relaxation matrices $\Lambda_{d,o}$, related to $\{\Theta^\dagger\Theta,G_S\}$, are given by
{
\begin{equation}
\Lambda_\zeta=\left(\begin{array}{ccccc}
2{G_\zeta}_{11}&{G_\zeta}_{12}&\dots&{G_\zeta}_{1\;N-1}&2{G_\zeta}_{1N}\\
{G_\zeta}_{21}&0&\dots&0&{G_\zeta}_{2N}\\
\vdots&\vdots&0&\vdots&\vdots\\
{G_\zeta}_{N-1\;1}&0&\dots&0&{G_\zeta}_{N-1\;N}\\
2{G_\zeta}_{N1}&{G_\zeta}_{N2}&\dots&{G_\zeta}_{N\; N-1}&2{G_\zeta}_{NN}
\end{array}
\right)
\end{equation}
with $\zeta=o,d$. Notice that these matrices contain an identically vanishing internal $(N-2)\times (N-2)$ square.

It can be proven that the steady state is unique \cite{dhahri,prosen} and reached exponentially with a relaxation time $T\sim N^3$ \cite{prosen}. 
The steady state correlation matrix $G_S^*$ obeys (\ref{eqfond3}) with the left-hand side set to zero.
In the vanishing square sector of $\Lambda_{d,o}$, thanks to the antisymmetry of the correlation matrix $G_S$, one derives from (\ref{eqfond3}) the space translation invariance of the matrix $G^*_d$: $(G^*_d)_{kl}=G^*_d(l-k)$.
It implies in particular that the steady state magnetization current,  
$\langle{\cal J}_k^m\rangle^*=2J{G^*_d}_{k\;k+1}\equiv 2 J \jmath^*$, 
defined through the quantum continuity equation 
$
\dot{\sigma^z_k}=i[H_S,\sigma^z_k]={\cal J}^m_{k-1}-{\cal J}^m_k
$,
is constant all along the chain. 
Using this translation invariance, one reduces the full set of steady equations to 
\begin{equation}\left\{\begin{array}{rl}
m_1^*-m^*=\gamma \jmath^*=m^*-m_N^*\\
m_L-m^*_1=\jmath^*/\gamma=m^*_N-m_R
\end{array}\right.
\label{eqsteady}
\end{equation}
with $\gamma=\frac{J_E^2}{2J}$ and where $m_k^*=\langle \sigma^z_k\rangle=-(G^*_o)_{kk}$ is the steady state magnetization at site $k$, which takes a constant value denoted $m^*$ $\forall k=2,...,N-1$.  One finds that all other correlation matrix elements are identically vanishing in the steady state.
From (\ref{eqsteady}) it appears that the four unknowns $\jmath^*, m^*,m_1^*$ and $m_N^*$ are functions of $\gamma$ and $m_{L,R}$ (that is on $\beta_{L,R}$ and $h$) only and consequently size-independent.
Solving (\ref{eqsteady}) one finally finds that the exact steady state properties are fully characterized by a stationary current
\begin{equation}
\jmath^*=\frac{\gamma}{1+\gamma^2}\frac{m_L-m_R}{2}
\end{equation}
and a flat magnetization profile
\begin{equation}
m^*_k=m^*=\frac{m_L+m_R}{2} \quad \forall k=2,...,N-1\\
\end{equation}
for the bulk spins and 
$ m^*_1=m^*+\gamma \jmath^*$, $ m^*_N=m^*-\gamma \jmath^*$
for the boundary sites. 
The size-independence of these quantities has to be related to the perfect ballistic nature of the elementary excitations transport properties.
One may notice that the boundary values are deviating from the flat profile by an amount which is proportional to the current value. However, in the large reservoir coupling limit, $\gamma\rightarrow \infty$, the magnetizations of the left and right boundary sites tend to the corresponding reservoir values. It is interesting to note that while the bulk magnetization profile is independent of the interaction strength ratio, $\gamma=\frac{J_E^2}{2J}$, the current show a non-monotonous behavior, see figure (\ref{fig2}), with a maximal current state for $\gamma=1$. One may explain this behavior by noticing that at small $\gamma$ the system is very weakly coupled to the reservoirs and it is very unlikely to inject a particle (or flip the boundary spin) at the boundary site, leading to a small current value. On the contrary for large $\gamma$ the coupling to the reservoirs is much larger than the chain coupling and it is very easy to flip the boundary spin, but hard to propagate this flip along the chain and this leads again to a small current. 
\begin{figure}
\centerline{
\includegraphics[width=6.cm,angle=0]{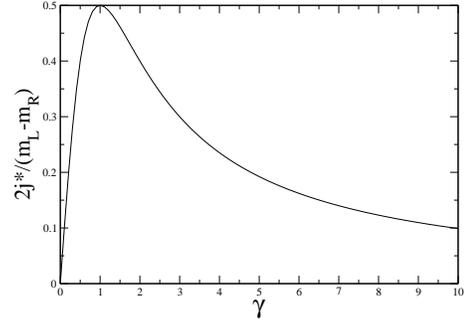}
}
\caption{\label{fig2} Rescaled stationary current as a function of the coupling ratio $\gamma=\frac{J_E^2}{2J}$.
}
\end{figure}

From the knowledge of the two-point correlation matrix one can deduce the steady state 
$\rho_S^*$ of the system, which will appear to be of generalized Gibbs form.
To show that, let us introduce Dirac Fermions: $\Gamma_k^1=c_k^++c_k$ and $\Gamma_k^2=i(c^+_k-c_k)$ and search for a quadratic form ${\cal Q}=\sum_{i,j}c^+_i A_{ij}c_j$ such that $\rho_S^*\sim e^{-{\cal Q}}$.
The coupling matrix $A$, giving $\cal Q$, is deduced from the correlation matrices $G_{d,o}^*$ or equivalently, using the Dirac Fermions representation, from ${\cal N}_{ij}=Tr\{ c_j^+c_i \rho_S^*\}$ thanks to the relation
$
A=\ln\left(\frac{1-\cal N}{\cal N}\right)
$ 
\cite{peschel,cheong}. 

In the large system size limit, after the diagonalization of the correlation matrix $\cal N$, one obtains from the previous relation
$A_{k+l,k}=(-1)^l A_{k,k+l}=\alpha_l$ where
$
\alpha_{l\neq 0}={\rm sgn}^l(\jmath^*)\frac{(i)^l}{l}\left[(-1)^l z^{l}(\frac{n^*}{\jmath^*})-z^{l}(\frac{1-n^*}{\jmath^*})\right]
$
with $z(x)=|x|-\sqrt{x^2-1}$ and 
$
\alpha_0=\ln\left(z(\frac{n^*}{\jmath^*})/z(\frac{1-n^*}{\jmath^*})\right)
$. Using that into $\cal Q$ one finally obtains the generalized Gibbs form
$
\rho_S^*\sim e^{-\alpha_0 Q_0/2-\sum_{l\neq 0} \alpha_l Q_l}
$
with $Q_{l}=\sum_j c^+_{j+l}c_j+(-1)^l c^+_jc_{j+l}$ a set of conserved quantities.

To interpret this result, and eventually extract an effective temperature characterizing the QNESS, consider first the undriven situation $\jmath^*=0$ for which $\beta_{L,R}=\beta$ ($n_{L,R}=n^*$). In that case, since $\alpha_l \propto {(\jmath^*)}^l$, all $\alpha_{l\neq 0}=0$ while  $\alpha_0$ reduces to $\ln\frac{1-n^*}{n^*}=\beta h$. Consequently the stationary state reached in our setup is described by the equilibrium Gibbs state $e^{-\beta H^0}$ with a free spin reference Hamiltonian $H^0=\frac{h}{2}M^z=\frac{h}{2}Q_0+const.$ and a temperature $\beta^{-1}$ set by the bath's temperature.  

To the lowest order in $\jmath^*$ the state reduces to the near-equilibrium form
$
\rho_S^*\sim e^{-\beta_{eff} H^0+\frac{\jmath^*}{2n^*(1-n^*)} J_1}
$
with $\beta_{eff}=\frac{1}{h}\ln\frac{1-n^*}{n^*}$ and $J_1=iQ_1$ the current operator with expectation $\langle J_1\rangle =\jmath^*$. 
At high temperatures $\beta_{eff}$ reduces to $\bar{\beta}=(\beta_L+\beta_R)/2$.

From this analysis, it appears that the identification $\beta_{eff}=\alpha_0/h$ is physically grounded.
For a finite current value one may use the symmetry property $\alpha_l(-\jmath^*)=(-1)^l\alpha_l(\jmath^*)$ to split $\sum_{l\neq0}\alpha_l Q_l$ into a current-like part
$
{\cal Y}=\sum_{l\ge 0} \alpha_{2l+1}Q_{2l+1}
$
, which is odd under boundary reflection, and a remaining  even part
$
{\cal K}= \sum_{l\ge 1} \alpha_{2l}Q_{2l}
$ \cite{activity}.
The steady-state is written then
$
\rho_S^*\sim e^{-\beta_{eff}H^0- {\cal Y}-{\cal K}}
$ 
with an effective inverse temperature $\beta_{eff}=\alpha_0/h$
that can be decomposed into 
$
\beta_{eff}=\beta_{conf}+\Delta({\jmath^*}^2)
$
where $\beta_{conf}=\frac{1}{h}\ln \frac{1-n^*}{n^*}$ is the configurational (level population)  part and $\Delta$ the current contribution. Since $\Delta \sim {\jmath^*}^2$ at small currents, the current contribution to the temperature does not show up in the linear regime.

In summary, we have obtained the exact QNESS of a finite XX chain in contact at both ends with repeated-interaction reservoirs. 
We have shown that in the steady state, the system is completely specified by two quantities, namely the magnetization profile (particle density) and the associated current. The flatness of the magnetization profile is related to the integrability of the model, leading to the violation of Fourier law since the system shows ideal conductivity. The QNESS is given by the generalized Gibbs state 
$
e^{-\beta_{eff}H^0-{\cal Y}-{\cal K}}
$
at inverse temperature $\beta_{eff}$ with respect to a reference Hamiltonian $H^0$. The many-body terms $\cal K$ and $\cal Y$, build on system conserved quantities, are
respectively symmetric and antisymmetric with respect to $\jmath^*$.

We would like to thank S. Attal for useful discussions.


\begin{thebibliography}{99}
\bibitem{remark0} The bath's spin states are completely characterised, through their Curie-Weiss magnetization $m_{L,R}=-\tanh(\beta_{L,R}h/2)$, by the temperatures $\beta^{-1}_{L,R}$.
\bibitem{remark1} The replacement of a reservoir spin that has already interacted with the system by a fresh one strongly reminds of the Boltzmann historical Stosszahlansatz in kinetical theory.
\bibitem{remark2} Without any rescaling, the limit $\tau\rightarrow 0$ leads trivially to the decoupled dynamics ${\cal L}(.)=i[T_S,.]$, where $T_S$ is the system interaction matrix, insuring a unitary evolution of the system.
\end{thebibliography}
\end{document}